\documentclass[a4paper,11pt]{article}
\pdfoutput=1 

\usepackage{jinstpub} 

\title{Challenging luminosity measurements at the Electron-Ion Collider}

\author{K. Piotrzkowski}

\affiliation{Université catholique de Louvain, Centre for Cosmology, Particle Physics and Phenomenology,\\Chemin du Cyclotron 2, 1348 Louvain-la-Neuve, Belgium}

\emailAdd{krzysztof.piotrzkowski@cern.ch}

\abstract{A precise determination of absolute luminosity, using the bremsstrahlung process, at the future Electron-Ion Collider (EIC) will be very demanding, and its three major challenges are discussed herein. First, the bremsstrahlung rate suppression due to the so-called beam size effect has to be well controlled. Secondly, the impact of huge synchrotron radiation fluxes should be mitigated. Thirdly, enormous bremsstrahlung event rates, in excess of 10~GHz, have to be coped with. A basic layout of the luminosity measurement setup at the EIC, addressing these issues, is proposed, including preliminary detector technology choices. Finally, the uncertainties of three proposed methods are also discussed.}

\keywords{Instrumentation for particle accelerators and storage rings - high energy, Calorimeters, Particle tracking detectors}

\arxivnumber{2106.08993} 

\begin{document}
\maketitle
\flushbottom

\section{Introduction}
\label{sec:intro}

Precise measurements of the electron-hadron cross sections are the corner stone of scientific program at the future Electron-Ion Collider (EIC), hence high demands on the EIC luminosity measurements – at least a 1\% accuracy is required for the absolute luminosity determination and only a $10^{-4}$ uncertainty is expected for the relative luminosity measurements, relevant for precise spin asymmetry studies at the EIC~\cite{YR}. 

It was demonstrated at HERA – the first electron-hadron collider – that the bremsstrahlung process can be successfully used to precisely measure the luminosity of high energy $ep$ collisions~\cite{hera1,hera2}. As it is discussed in the following, such a technique can be also used at the EIC, but it poses a major challenge, and a wide range of the electron beam energies and a large variety of hadron species, from protons to gold nuclei, will only increase that challenge. It should be noted that the sensitivity of the bremsstrahlung process to the beam polarizations and its impact on the relative luminosity measurements are beyond the scope of this paper.

This article is structured as follows. In section~\ref{sec2}, unique properties of high energy bremsstrahlung are discussed; section~\ref{sec3} gives a description of  the forward photon detector setup and of the principal bremsstrahlung background; section~\ref{sec4} gives an account of conditions for detection of very forward scattered electrons, and in sections~\ref{sec5} and~\ref{sec6}, respectively, the proposed detector technologies and major contributions to uncertainties of the luminosity measurement are discussed. The paper is concluded with a summary in section~\ref{sec7}, and in appendix~\ref{app} supplementary details are given regarding the discussed experimental methods.

\section{\label{sec2}Bremsstrahlung at the EIC}

\begin{figure}[htbp]
\includegraphics[width=0.5\textwidth]{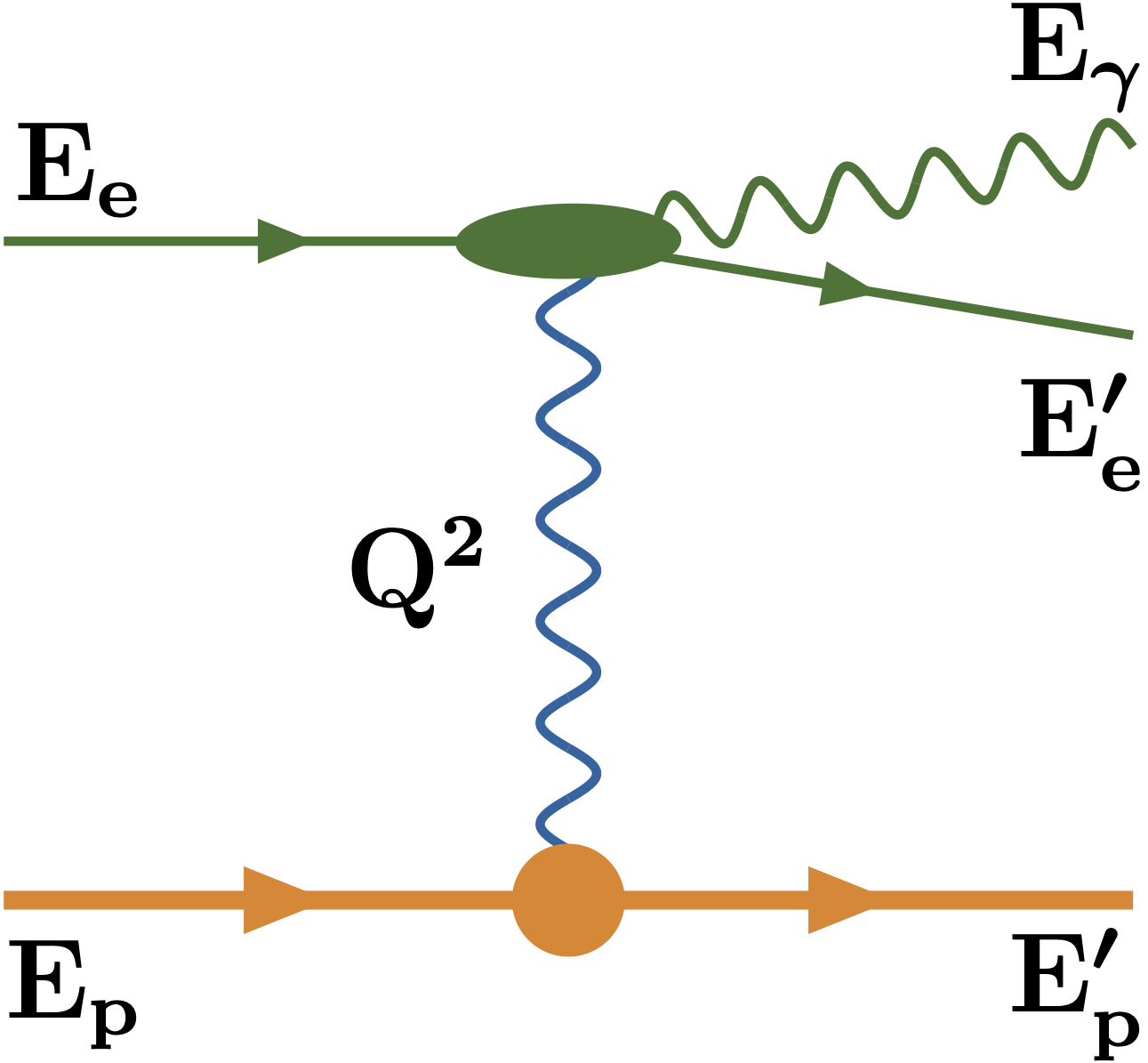}
\caption{\label{fig_1} Feynman diagram for the electron-proton brems\-strahlung. Respective energy variables of incoming and outgoing particles are shown as well as the virtuality of exchanged photon $Q^2=-q^2=-(P_p-P_p^\prime)^2$, where $P_p$ and $P_p^\prime$ are four-momenta of the incoming and outgoing protons, respectively.}
\end{figure}

The high energy bremsstrahlung process is so unique due to extremely small momentum transfers between the radiating electron and the other charged particle, a proton in case of the electron-proton collisions, as in figure~\ref{fig_1}. The typical bremsstrahlung photon emission angle is very small, $\approx m_e/E_e$, but it is kinematically allowed that both incoming particles experience no angular scattering, that is, continue moving exactly along their initial directions, while the photon is emitted exactly in the direction of the incident electron momentum. It is precisely this configuration which results in the smallest virtuality $Q^2$ of the exchanged photon (see figure~\ref{fig_1} for definitions of variables):
\begin{equation}
|Q|_\mathrm{min}\simeq m_e^2 m_p E_\gamma /(E_e^\prime s_{ep}) ,
\label{eq_1}
\end{equation}
where $s_{ep}$ is the $ep$ center-of-mass energy squared, and $m_e, m_p$ are the electron and proton mass, respectively.\\

One should note that the higher the beam energies are lower the minimal virtuality becomes. For example, at the EIC, for $E_e=18$~GeV, $E_p = 275$~GeV and $E_\gamma =1$ GeV, one gets the minimal {\it longitudinal} momentum transfer in the proton rest frame, $|q|_\mathrm{min}/\mathrm{c} \simeq 0.0007$ eV/c. Therefore, the corresponding (kinetic) energy transfer $|q_\mathrm{min}|^2/2m_p\mathrm{c}^2\approx 3\times10^{-16}$~eV, and for all the practical purposes one can assume that $E_e^\prime+E_\gamma=E_e$ exactly.

The bremsstrahlung differential cross section, at a fixed photon energy and with unintegrated scattering angles, is proportional to $1/Q^4$ (due to the propagator of exchanged photon), therefore the photon virtualities close to $Q^2_\mathrm{min}$ dominate, and to a very good approximation $Q^2 = Q^2_\mathrm{min} + q_\perp^2$, where $q_\perp^2$ is the transverse component of four-momentum transfer. If one analyses this process in the impact parameter space \cite{Kotkin:1992bj}, these very small values of $|q_\perp|$ correspond to very large values of the impact parameter $b = \hbar/|q_\perp|$. That, in turn, explains why the original Bethe-Heitler cross section calculations in the Born approximation are so precise, despite neglecting the size of the proton and its spin. 

The impact parameter in high energy bremsstrahlung reaches truly macroscopic values – the above example for the EIC corresponds to $b$ of about 0.3~mm. In consequence, when {\it both} colliding beams are strongly focused, this may lead to significant breaking of the basic relation: 
\begin{equation}
R=L\sigma,
\label{eq_2}
\end{equation}
where $R$ is the rate of events of a given process with the cross section $\sigma$ and $L$ is the factor called luminosity which depends on fluxes of colliding particles. In derivation of eq. \ref{eq_2} it is assumed that all these particles can be represented by plane waves. This approximation works well as long as the transverse beam sizes are much bigger than the typical impact parameters involved, but just that condition is not fulfilled in bremsstrahlung at high energy colliders.

The effective suppression of bremsstrahlung due to finite beam sizes was observed in high energy electron-proton collisions at HERA \cite{hera-bse}. In spite of lower beam energies, this beam size effect (BSE) will be stronger at the EIC due to its much smaller lateral beam sizes \cite{bse-paper}. Somewhat paradoxically, the eq. \ref{eq_2}, to be applied for the luminosity determination at the EIC, is violated just in case of bremsstrahlung to be used for that purpose. However, it was shown in ref. \cite{bse-paper} that the large impact parameter nature of bremsstrahlung can be thoroughly  studied at the EIC using the lateral beam scans.
\section{\label{sec3}Measurements of zero-degree photons at the EIC}
Already very first measurements of bremsstrahlung at HERA showed that, thanks to its large cross section, all backgrounds were small and there was no need to detect the scattered electrons to obtain a clean bremsstrahlung signal \cite{hera1}. This greatly simplified the luminosity determination as, apart from the photon energy selection, only the high geometrical acceptance of bremsstrahlung photons had to be taken into account, resulting in a 1\% accuracy of absolute luminosity measurements at HERA~I~\cite{hera2}. At HERA~II, however, the experimental conditions were much worse for the luminosity measurements due to a much stronger direct synchrotron radiation background. This made such direct measurements of bremsstrahlung photons much more difficult and less precise. To overcome that problem, and to be at the same time less sensitive to the event pileup, the ZEUS collaboration also introduced another technique, utilizing high energy $e^+e^-$ pairs from the bremsstrahlung photon conversions in the beampipe exit window -- still, only a 2\% absolute luminosity precision was achieved~\cite{hera3}. The H1 collaboration using solely direct measurements of  bremsstrahlung photons, corrected for the event pileup, achieved an absolute precision of about 3\%~\cite{hera4}, which was slightly improved by use of the elastic QED Compton events for an overall luminosity normalization~\cite{hera5}.
\begin{figure}[htbp]
\includegraphics[width=1.04\textwidth]{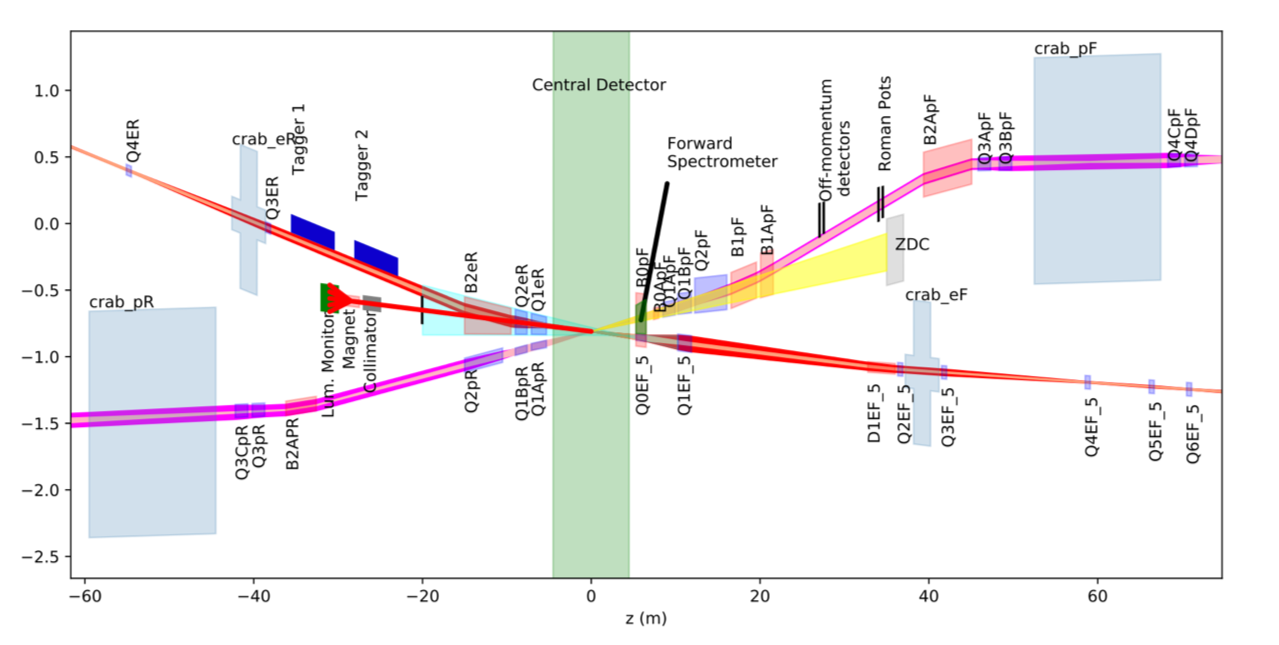}
\caption{\label{fig_2} Schematic layout of the EIC interaction region with a beam crossing angle of 25 mrad. Note the length scales for the horizontal and vertical axis are very different. It is figure 3.3 from ref. \cite{cdr}.}
\end{figure}

A preliminary layout of the interaction region at the EIC is shown in figure \ref{fig_2}, including  sketches of the luminosity detectors \cite{cdr}. The bremsstrahlung photon exit window is located about 20~m from the interaction point (IP), and only three electron beamline elements are present in-between -- a quadrupole doublet, Q1eR and Q2eR, and a dipole B2eR. The EIC design ensures such internal apertures of these magnets that all photons emitted from the IP at the angle $< 1$ mrad, with respect to the nominal electron beam axis there, will reach the exit window (see Table 8.1 in \cite{cdr}).

The synchrotron radiation (SR) has similar angular distribution as bremsstrahlung\footnote{SR is also known as \it magnetobremsstrahlung radiation.} -- it is emitted at very small angles ($\approx m_e/E_e$) with respect to the instantaneous direction of motion of a beam electron. Since the electron beam goes through the centers of both Q1eR and Q2eR, therefore B2eR is the main source of the direct SR hitting the photon exit window. This dipole deflects the electron beam by about 18~mrad to separate it further from the hadron beamline, and at the same time serves as an analyzing magnet for forward scattered electrons.
For 10 and 18~GeV electron beams this would be a source of a huge SR flux -- the power reaching the exit window would exceed 4~kW. It was therefore proposed to split this dipole magnet in two parts, where the first one, relevant for the luminosity detectors, has about 4 times weaker field than B2eR.

\begin{table}[htbp]
\centering
\caption{\label{tab} Bethe-Heitler $ep$ bremsstrahlung cross sections in mb (and the corresponding event rates in GHz, for the nominal EIC luminosities), for various beam energies in GeV and three selection criteria.}
\smallskip
\begin{tabular}{|lr||c|c|c|}
\hline
$E_e$ & $E_p$ & $E_\gamma/E_e > 0.01$ & $1> E_\gamma/E_e > 0.7$ & $0.4> E_\gamma/E_e > 0.1$ \\
\hline
\hline
18 & 275 & 237 (0.36) & 11.6 (0.018) & 65.2 (0.10) \\
10 & 275 & 230 (2.3) & 11.1 (0.11) & 63.2 (0.63) \\
5 & 100 & 209 (0.77) & 9.81 (0.036)& 57.1 (0.21)\\
\hline
\end{tabular}
\end{table}
Minimizing the direct SR flux is of fundamental importance for the luminosity measurement at the EIC as it affects both the converted and direct bremsstrahlung photon measurements (see section~\ref{sec5}). The third challenge, after the BSE and SR ones, is due to enormous bremsstrahlung event rates. In table \ref{tab} the $ep$ bremsstrahlung cross sections and the corresponding event rates are given for three examples of beam energies, and for three selection criteria: first, corresponding to the direct photon measurement; second, to the selection of high energy photons; third, to the  bremsstrahlung photons tagged by forward electron detectors. One should note, as this will affect the event pileup, that for 18 GeV electron beams the bunch crossing rate will be only 25 MHz, due to the SR power limitations, whereas for 5 and 10 GeV ones -- 100 MHz. As a result, for the first selection in table \ref{tab}, the average number of bremsstrahlung events per bunch crossing will be about 23 and 14, for 10 and 18 GeV beams, respectively\footnote{In the end, the nominal total energy deposited per bunch crossing will be slightly bigger for 18 GeV beams.}. The rates will be even higher for the electron collisions with heavy ions, as the bremsstrahlung cross sections scales with the colliding particle charge squared. For example, in case of $e$Au ($Z=79$) collisions at 10 GeV the expected bremsstrahlung event pileup is almost 350, corresponding to the event rate above 30~GHz. These extreme event rates will result in significant power deposited in the middle of PCALf -- of about 15~W for the $e$Au case, and about 1~W for the $ep$ one. 

\begin{figure}[htbp]
\includegraphics[width=0.95\textwidth]{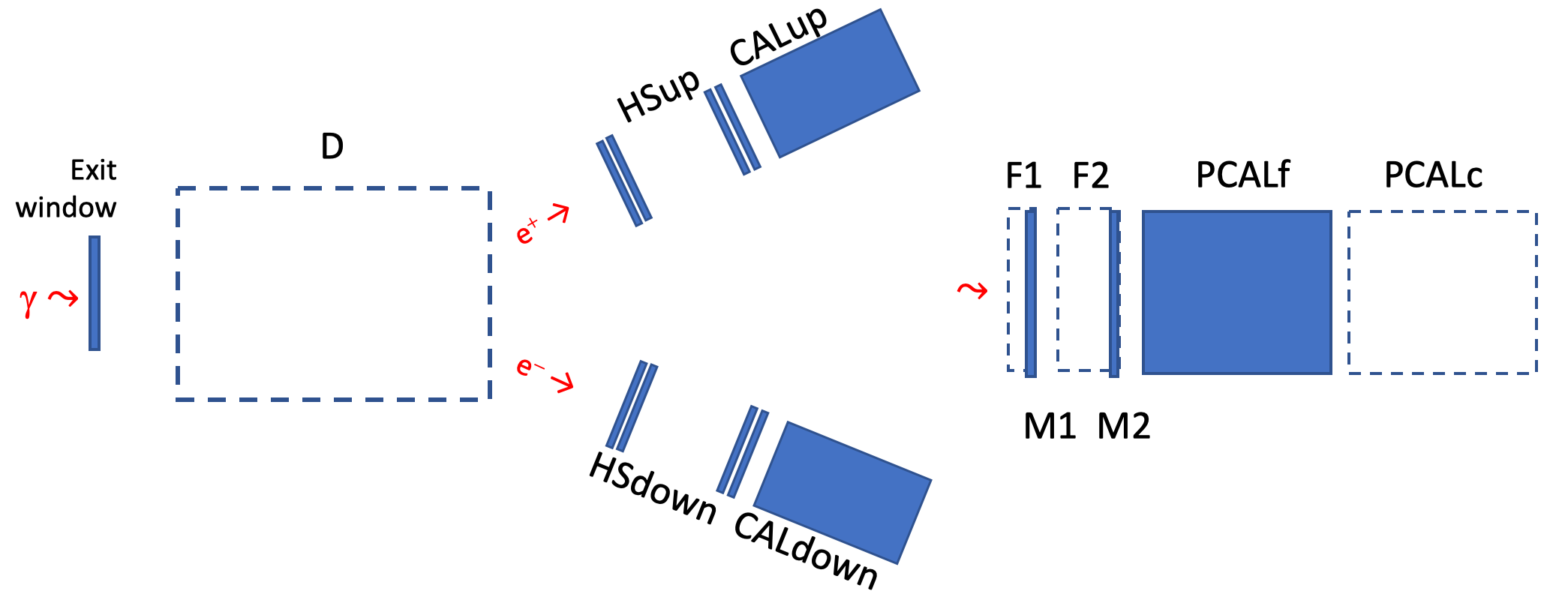}
\caption{\label{fig_3} Conceptual layout of the bremsstrahlung photon detection setup (side view), composed of two parts. The spectrometer part will measure the $e^+e^-$ pairs from the  photon conversion in the exit window, and consists of a small dipole magnet D with adjustable horizontal field, two calorimeters, CALup and CALdown, and two hodoscopes, HSup and HSdown. The direct part will measure the unconverted photons and includes two movable calorimeters, PCALf and PCALc, for use at high and low luminosity, respectively, and two different SR filters, F1 and F2, which can be remotely inserted, and are instrumented with the SR monitors, M1 and M2.}
\end{figure}

In figure \ref{fig_3} a conceptual layout of the proposed detectors of bremsstrahlung photons at the EIC is shown. It consists of the spectrometer part for the $e^+e^-$ detection, which includes two calorimeters and two hodoscopes, and of the direct part with two movable calorimeters for detection of unconverted bremsstrahlung photons, PCALf and PCALc, for use at high and low luminosity, respectively. In front of these two calorimeters there are two different SR filters, which can be remotely inserted, and are both instrumented with SR monitors.

\section{\label{sec4}Very low-$Q^2$ electrons at the EIC}
As indicated in figure \ref{fig_2}, there is a place foreseen in the electron beamline for installation of Taggers -- the electron detectors, which will register both the bremsstrahlung electrons as well as will tag photoproduction events, corresponding to generic very low-$Q^2$ scattering at the EIC. The detectors will be placed around 30~m from the IP, behind the B2eR dipole, so the scattered electrons which lost sufficient energy will be deflected enough to reach them\footnote{For example, an electron which lost 10\% of its initial energy will be deflected in B2eR by 20~mrad, in place of 18~mrad, resulting in lateral displacement of this electron by 40~mm, with respect to the electron beam axis, at a 20~m distance behind the dipole center.}.

\begin{figure}[htbp]
\includegraphics[width=0.90\textwidth]{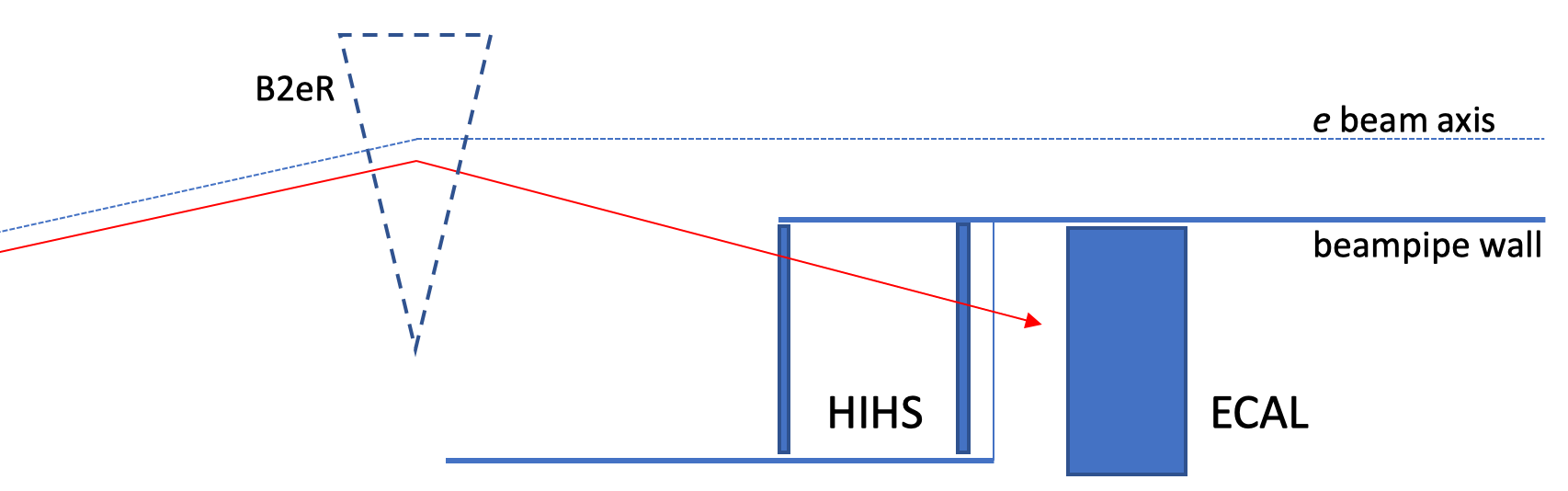}
\caption{\label{fig_4} Conceptual layout of the forward electron detection setup (top view). It consists of a calorimeter ECAL and a high resolution hodoscope HIHS in the electron beam primary vacuum.}
\end{figure}

In figure \ref{fig_4} a conceptual layout of the proposed detectors of very low-$Q^2$ electrons at the EIC is shown. It consists of a calorimeter behind an electron exit window (ECAL) and a high resolution hodoscope (HIHS) inserted into the beam primary vacuum, to avoid the electron multiple scattering in the exit window.

Regarding the luminosity measurements only, ECAL is more or less sufficient, provided its good energy and spatial resolutions, as it will serve there primarily to cross check the bremsstrahlung photon energy scales as well the acceptance corrections, including the photon conversion factor.

However, this is highly insufficient for the  photoproduction tagging at the EIC because of a very high event pileup due to bremsstrahlung -- if one assumes the tagging range equal to the third selection in table \ref{tab} (that is $0.4> E_\gamma/E_e > 0.1$) then for 10~GeV electron beams the average number of simultaneous bremsstrahlung electrons in this range is 6.3 for the $ep$ collisions, and almost 100 for the $e$Au ones, at the nominal luminosities. Clearly, even highly segmented calorimeters cannot cope with it and a dedicated hodoscope is needed.

Such a hodoscope should measure very precisely, for each track, the positions and angles in horizontal and vertical planes. Solely from these measurements four initial parameters can be in principle calculated -- the two electron momentum angles at the IP, as well as the event horizontal vertex position and the electron energy. Note that the vertical vertex position smearing is very small at the EIC, usually $\sigma_y < 10\,\mu$m, so the vertical vertex position can be assumed fixed here. However, the calorimetric energy measurement is still necessary to perform kinematic fitting and improve the reconstruction, in particular for events with exceptionally small angles at the IP, in the vertical plane. Then, to properly associate a given very forward scattered electron with the photoproduction event measured by Central Detector, one can use, apart from the event kinematics matching, the consistency of the two completely independently reconstructed horizontal positions of the event vertex.

In such a case, the luminosity measurement would profit from very precisely reconstructed bremsstrahlung electron energies, as one could use these events to cross calibrate the photon detectors, yet with higher precision. The electron energy reconstruction with such a hodoscope can be very strongly verified by analyses of the exclusive production of charged lepton pairs at the EIC. This would then effectively link the bremsstrahlung photon energy scales with the very precisely controlled energy scale of the central tracking detectors\footnote{While running at low event pileup such tagged exclusive events can be measured with ECAL only, though with less precision.}. One should note here that the main physics motivation for tagging photoproduction is to improve studies of the exclusive photoproduction of vector mesons, decaying leptonically, which themselves provide this cross calibration opportunity.
\section{\label{sec5}Detector challenges and possible technology choices}
The choice of the detector technology in case of the calorimeters is dictated mostly by the corresponding bremsstrahlung event rates. For the direct photon measurements at the nominal luminosity an extremely radiation hard option has to be used. For 10~GeV electron and 275~GeV proton beams, the number of detected bremsstrahlung photons for the integrated luminosity of 100~fb$^{-1}$ is about $2.3\times10^{16}$ (see table \ref{tab}), what results, for the average photon energy of about 2.5~GeV, in the total absorbed energy of about $6\times10^{25}$~eV. Assuming use of sampling calorimeters with a sampling fraction of 0.01 and that 1\% of that sampled energy is deposited per 1~g of the active material in the most irradiated region, one obtains a rather conservatively approximated dose of 1~MGy there. This prevents any use of scintillators but the fused silica are expected to stand rather well such irradiations \cite{silica}. The spectrometer calorimeters and ECAL will face less than 1\% of that irradiation dose, allowing to use modern rad-hard scintillating materials, studied and developed for the HL-LHC, for example~\cite{scint}.

Therefore, as a starting point for further studies it is proposed to assume for PCALf a tungsten spaghetti calorimeter with fused silica fibers. The Cerenkov light is read out separately from each fiber with a silicon photomultiplier (SiPM), where the expected number of fibers is of the order of 2000. The SiPM signals are sampled at 100 MHz and the single channel occupancy approach 1 only for the most extreme case of $e$Au collisions.

In addition, the synchrotron radiation flux should be monitored using the dedicated detectors behind the SR filters (M1 and M2). It is proposed to use for this purpose the same fused silica material as in PCALf. Finally, a dedicated detector of the coherent bremsstrahlung should be installed to improve the crucial beam diagnostics at the IP~\cite{bse-paper}.

The other four calorimeters (that is PCALc, CALup, CALdown and ECAL) should be built using the same technology, as tungsten spaghetti calorimeters with scintillating fibers, read out by SiPMs, for example. The three hodoscopes are essential not only for a better event reconstruction but also to allow for a direct and precise on-line monitoring of the calorimeter uniformities and linearities -- which is critical given huge irradiaton doses involved and the inevitable gradual material damage.

The spectrometer hodoscopes, HSup and HSdown, will not face very large event rates, and as only the vertical track position has to be determined, just several simple planes of about 1~mm square, straight scintillating fibers read out by SiPMs are proposed for that purpose. To ensure an optimal performance of the spectrometer detectors the photon exit window should be made as thin as possible. A 10~mm thick aluminium window (equivalent to 11\% $X_0$)  significantly spoils the photon energy reconstruction (see figure 11.114 in ref. \cite{YR}) -- therefore, reduction to a 5~mm thickness is highly desirable. A thin exit window is particularly important for the spectrometer hodoscopes as the multiple scattering of the electrons and positrons spoils in addition the proper spectrometer momentum measurement. For example, a 5~GeV electron will experience in a 5~mm thick aluminum window an average angular dispersion of about 0.4~mrad, in the vertical plane. Assuming the corresponding bending angle in the spectrometer dipole of 40~mrad one arrives to a 1\% limit on the momentum resolution limit due to the multiple scattering. Such an angular dispersion defines then what the track angular resolutions of the HSup and HSdown hodoscope should be. The aforementioned two stacks of planes of 1~mm fibers separated by 1--2~m should very well provide that.

The experimental conditions are very different for HIHS -- the track rates and densities can be very high in this case, therefore highly pixelized and thin semiconductor detectors seem to be mandatory. Finally, huge event rates will require very intensive near-detector signal processing. This should be studied as soon as the detectors' characteristics will be better defined.
\section{\label{sec6}Determination of absolute luminosity at the EIC}
First and foremost, the beam size effect should be properly investigated using PCALc and the lateral beam scans as proposed in ref.~\cite{bse-paper}. It should be noted that the BSE observations at HERA have verified the theoretical predictions at a 30\% level only, and in a limited photon energy range~\cite{hera-bse}. Such dedicated studies at the EIC will allow to properly assess the errors of the BSE corrections, due to uncertainties in the electron and hadron beam sizes, the possible lateral beam displacements, and other parameters of the BSE modelling in bremsstrahlung.

In addition, such special bremsstrahlung measurements at low event pileup should be repeated on a regular basis, as the colliding beam "geometries" will change in time, whereas the detector understanding will hopefully improve. Moreover, these special calibration runs will produce very complete and clean bremsstrahlung data samples from all the relevant detectors to allow for making regularly the crucial, data driven self- and cross-calibrations.

Two largely independent luminosity measurements should be performed simultaneously at the EIC, as they have very complementary features -- one is based on the direct photon measurement using PCALf and the other uses the spectrometer detectors to count number of converted photons in a certain energy window. This guarantees an improved accuracy of the luminosity measurement in strongly varying experimental conditions -- as in the case of proton vs. heavy ion beams, or high vs. low energy beams, or finally, high vs. low luminosity running.

For example, for the 10 GeV electron beam colliding with the 275 GeV proton beam at the nominal luminosity, the average bremsstrahlung energy per bunch crossing deposited in PCALf is about 60~GeV and is simply a measure of the instantaneous EIC luminosity per bunch crossing. In such a way, one can very directly determine the EIC absolute luminosity, as the bremsstrahlung photon geometrical acceptance should be very high, hence the corresponding correction very small and well controlled. One should note, that for the $e$Au collisions this average energy will reach 1~TeV, at the nominal luminosity. This luminosity measurement method is in principle very simple but is a subject to significant BSE corrections, and for 18~GeV electron beams might suffer from non-negligible corrections due to the strong SR background.

In contrast, the spectrometer method is by construction not sensitive to the SR backgrounds, and is less sensitive to the BSE, but has a complex acceptance for the $e^+e^-$ pairs, requires an excellent evaluation of the photon conversion factor and in addition the large event pileup for the electron-heavy ion collisions is not straightforward to correct for in this case. 

Using two largely independent methods will significantly increase the overall robustness of the luminosity measurement at the EIC and will automatically provide a stringent test of the error estimates. The two nominal methods of the EIC luminosity determination, as well as the special, reference one using PCALc, are discussed in more detail in the appendix~\ref{app}, including their uncertainties and complementarity. Finally, to maintain the continuous control of the detector calibrations, the several electron pilot bunches (i.e. with the opposite hadron bunches not filled) should always be kept to allow for a "parasitic" accumulation of indispensable $e$-$gas$ bremsstrahlung data, with no event pileup \cite{hera-bse,phd}. 
\section{\label{sec7}Summary}
The challenges for precise determination of the absolute luminosity at the EIC are huge, much bigger than at HERA. This is why it is proposed here to significantly extend the dedicated instrumentation to well control all the relevant systematic biases and to maximally exploit powerful, data-driven and high statistics calibration techniques. 

Very precise measurements of the relative luminosity at the EIC pose yet another big challenge, but are not discussed in this article and need separate, dedicated studies. 

Achieving a precise luminosity determination is usually a very long and challenging task by itself -- the most striking example was provided by the LEP saga of its ultra precise luminosity determination, which seems to end only recently, 25 years after stopping LEP~I~\cite{lep}.  
\appendix
\section{\label{app}Methods of the luminosity measurement at the EIC and their uncertainties} 

First method of the luminosity measurement at the EIC is based on the bremsstrahlung photon energy flow as measured with PCALf:
\begin{equation*}
 f \langle E_\gamma \rangle = L A_\gamma (1-C_\gamma^*)\int_{E_{\gamma,min}} E_\gamma d\sigma ,
\end{equation*}
where $f\simeq100$~MHz is the bunch crossing rate, $\langle E_\gamma \rangle$ is the measured average sum of photon energies, in a single bunch crossing, of $e$-$ion$ bremsstrahlung (i.e. after subtracting contributions from the $e$-$gas$ bremsstrahlung, and the SR), $L$ is the EIC luminosity, the geometrical acceptance $A_\gamma> 0.99$ is the probability of a bremsstrahlung photon to reach the exit window, and $C_\gamma^* = C_\gamma - \epsilon$, where $C_\gamma$ is the photon conversion factor and $\epsilon\simeq 9C_\gamma^2/14$ is an average fraction of the incident photon energy detected via the "secondary" bremsstrahlung of $e^+e^-$ in the exit window; for a 5(10)~mm thick aluminium exit window $1-C_\gamma^* = 0.961(0.924)$. By construction, this method is insensitive to the photon event pileup and its main error is due to the uncertainty of $E_{\gamma,min}$ -- the effective energy threshold which describes the effect of the PCALf energy resolution, including its non-linearities and non-uniformities. The $E_{\gamma,min}$ value will be precisely calibrated using mostly the high statistics samples of bremsstrahlung events when the scattered electron is also measured. It should be noted that at the nominal EIC running conditions, significant corrections due to the BSE will need to be taken into account while calculating the above integral. 

Second method of the luminosity measurement is based on counting the selected events, using the spectrometer system, when the bremsstrahlung photons convert in the exit window into $e^+e^-$ pairs:
\begin{equation*}
R = L A_\gamma C_\gamma A_{sel} C_{pu} \Delta\sigma,
\end{equation*}
where $R$ is the measured event rate of $e$-$ion$ bremsstrahlung (i.e. after subtracting contributions from the $e$-$gas$ bremsstrahlung), $A_{sel}$ is the acceptance of $e^+e^-$ pairs for given selection cuts, and includes effects of the resolution of spectrometer detectors, $C_{pu}$ is the correction for event pileup, and a priori can be larger or smaller than 1, and $\Delta\sigma$ is the corresponding bremsstrahlung cross-section which is expected to have small BSE corrections, as the selection will favor high energy photons. The dominant errors in this method are due to the uncertainties on $C_\gamma$ and $A_{sel}$ -- both can be calibrated using the aforementioned special data samples\footnote{However, smaller the $C_\gamma$ factor is, which is profitable both for the $e^+e^-$ reconstruction and $C_{pu}$,  more difficult it is to control experimentally.}. The $C_{pu}$ factor can be kept close to 1, even for the $e$Au collisions, as long as the HSup and HSdown hodoscopes can be used to strongly reduce ambiguities in pairing electrons and positrons.

Third method is simply based on counting (with the dipole field off, or on) all bremsstrahlung photons above certain high energy threshold, using PCALc:
\begin{equation*}
R = L A_\gamma C_{pu} \sigma_{thr},
\end{equation*}
where $C_{pu}>1$, and the cross-section $\sigma_{thr}$ accounts for small BSE corrections as well as for the energy threshold, including the PCALc resolution effects. Thanks to its high precision this can serve as the reference luminosity measurement during special runs at small instantaneous luminosity, when $C_{pu}$ is very close to 1 and the detector can be in addition self-calibrated using the spectrum endpoint at the electron beam energy. If the spectrometer dipole field is on, then one has to include the $1-C_\gamma^*$ factor and can on-the-fly cross calibrate the second method.

\acknowledgments

Many thanks to Mariusz Przybycie\'n for his careful reading of the manuscript and helpful comments.

%

\end{document}